\documentstyle[12pt]{article}
\begin{document}
\title{Influence of Compressibility on Scaling Regimes of
Strongly Anisotropic Fully Developed Turbulence}
\author{
N. V. Antonov$^{1}$, M. Hnatich$^{2}$, M. Yu. Nalimov$^{1}$\\
$^{1}$ Department of Theoretical Physics, St~Petersburg University,\\
Uljanovskaja 1, St~Petersburg---Petrodvorez, 198904 Russia,\\
$^{2}$ Institute of Experimental Physics, Slovak Academy of Sciences,\\
Watsonova 47, 04011 Kosice, Slovakia}
\maketitle
\date{}

\begin{abstract}
Statistical model of  strongly anisotropic fully developed
turbulence of the weakly compressible fluid is considered by means
of the field theoretic renormalization group.
The corrections due to compressibility to the infrared form of
the kinetic energy
spectrum have been calculated in the leading order in Mach number
expansion.
Furthermore, in this approximation
the validity of the Kolmogorov hypothesis on the  independence
of dissipation length
of velocity correlation functions in the inertial range has been proved.
\end{abstract}
\bigskip


\section{Introduction} \label {sec:1}

One of the oldest open problems in theoretical physics is that of
describing fully developed turbulence on the basis of a microscopic
model. The latter is usually taken to be the stochastic Navier-Stokes
equation
subject to an external random force
which mimics the injection of energy
by the large-scale modes, see, e.g. Ref. \cite{Monin}. The aim of the
microscopic theory is
to verify the basic principles of the celebrated Kolmogorov-Obukhov
phenomenological
theory \cite{Kolmogorov},  study deviations from this theory
and find the dependence
of various Green functions (velocity correlation
and response functions) on the times, distances,
external (integral) and internal (viscous) turbulence scales. Most
results are obtained  within the framework of numerous
semiphenomenological models which cannot be considered to be the basis
for  construction of a regular expansion in a certain small
(at least formal) parameter, see  Ref. \cite{Monin}.

One of the exceptions is provided by the renormalization group (RG) method
earlier successfully applied in the theory of critical behavior to explain
the origin of critical scaling and  calculate universal quantities
(critical dimensions and scaling functions) in the form of $\epsilon$
expansions, see Ref.\cite{Zinn}.

The RG was applied to the stochastic Navier-Stokes equation in Refs.
\cite{Nelson,Dominicis,Pismak,Frish}. For the isotropic homogeneous
turbulence of  incompressible viscous fluid it allows one to prove the
existence of
the infrared (IR) scale invariance with exactly known ``Kolmogorov''
scaling dimensions and the independence of the correlation functions
of the viscous
scale (the second Kolmogorov hypothesis), and  calculate a number of
principal constants in  a reasonable agreement with the experiment.
The detailed exposition of the RG theory of turbulence and the bibliography
can be found in the review paper \cite{UFN}.

As the model of isotropic incompressible fluid provides only
a simplified description of real turbulent flows, it
is interesting to generalize the model by taking into account
anisotropy, compressibility, inhomogeneity, real geometry, and so on.
In particular, in a number of papers the turbulence with the weak
\cite{Barton,Denis,Kim} and strong \cite{Busa} uniaxial anisotropy
has been studied. It was shown that, in the three-dimensional space,
the IR scaling regime characteristic of the isotropic case  survived
also if the anisotropy was included (in the language of the RG,
this means that the corresponding fixed point remains IR stable).

In Refs. \cite{Yakhot,Loran,Nalimov,Udalov}, the isotropic turbulence of
 compressible fluid was considered. The main difficulty
is that the corresponding field theoretic model is not multiplicatively
renormalizable, so that the RG technique is not directly applicable
to it (for this reason, the results obtained in Ref. \cite{Yakhot}
cannot be considered reliable, see the discussion in Refs.
\cite{Loran,Udalov}).

In Ref. \cite{Loran}, the problem of non-renormalizability was
solved  in the frame of expansion procedure in small Mach number
${\bf Ma}=v_c/c$
(where $v_c$ is the characteristic mean-square velocity and $c$ is
the speed of sound).
In the first nontrivial order $({\bf Ma})^{2}$
the problem was reduced to the calculation of scaling
dimensions of certain nonlocal composite fields (composite operators in the language
of field theory), constructed from the fields of
the renormalizable model of incompressible fluid.

Calculation of the scaling dimensions of composite operators is quite
a cumbersome task.
As a rule, their  renormalization involves their mixing with each other,
and in order to find the scaling dimension of a given operator,
one has to consider the entire
family of  operators that admix to it in the renormalization procedure.
The use of functional Schwinger equations and Ward identities, which
express the Galilean symmetry of the model, simplifies the problem and
in many cases allows us to find the dimensions exactly, see Refs.
\cite{Pismak,UFN,Hnatic,55}. Using this technique for isotropic
turbulence, the authors of
\cite{Loran} have calculated all the relevant scaling dimensions and,
with the aid of these results, proved the validity of the second Kolmogorov
hypothesis (independence of the velocity correlation function of
the viscosity) in the leading
order of $ ({\bf Ma})^{2}$. This is in agreement with the result obtained
previously
in \cite{Lvov} within the approach based on the self-consistent equations.
In Ref. \cite{Nalimov}, this proof was generalized
to all orders of the formal expansion in $ ({\bf Ma})^{2}$.

It should be stressed that the stability of the Kolmogorov fixed point in
the presence of anisotropy is obviously not {\it a priori}: the analysis
of the $d$-dimensional case shows that the stability is violated for
$d<2.68$ \cite{Kim,Busa} (the two-dimensional case requires special care,
see Ref. \cite{Runov}).
The stability of the Kolmogorov  regime is also destroyed for the
anisotropic magnetohydrodynamic turbulence \cite{Denis} and for the
strongly compressible fluid \cite{Udalov}.

In this paper, we study the effect of compressibility in the first
nontrivial order
of the expansion in the small $({\bf Ma})^{2}$
within the framework of a more realistic
model of the uniaxial anisotropic turbulence. The anisotropy is
not supposed to be small. Like in the isotropic case \cite{Loran},
the problem is reduced to the calculation of the scaling dimensions
of a class of nonlocal composite operators in the model of
incompressible strongly anisotropic turbulence considered in
\cite{Busa}. However, the
set of  relevant operators in this case
is much wider than in Ref.\cite{Loran}. Using the technique developed
in \cite{Pismak,UFN,Hnatic,55} and the results obtained in \cite{Busa},
we have found all the scaling dimensions exactly. The main result of
the paper is the substantiation of the validity of the  second
Kolmogorov hypothesis  mentioned above, for strongly anisotropic,
weakly compressible developed turbulence in the first nontrivial
order in the  Mach number.

\section{The model}  \label {sec:2}

In the stochastic theory of fully developed turbulence, the motion of
a viscous fluid is described by the Navier-Stokes equation
\begin{equation}
 \varrho [\partial_t v_i + v_j \partial_j v_i]=
\nu_0 \Delta  v_i + \nu_0'\partial_i \partial_j v_j
 -  \partial_i {\cal P} + f_i,
\label{NS}
\end{equation}
the continuity equation
\begin{equation}
\partial _t \varrho +  \partial_j(\varrho  v_j)=0,
\label{NR}
\end{equation}
and the equation of state
$ {\cal P} = {\cal P} (\varrho)$.
Here $\partial_t\equiv \partial /\partial t$,
$\partial_i\equiv \partial /\partial x_i$,
$v_i({\bf x},t)$ are the coordinates of the velocity field,
$\varrho({\bf x},t) $ is the density of the fluid,
$ {\cal P} ({\bf x},t)$ is the pressure, and
$\nu_0$ and $\nu_0'$ are the molecular viscosity coefficients.
Here and henceforth, summation over repeated indices is implied.

Following the tradition of stochastic models of turbulence, the
randomness in Eq. (\ref{NS}) is introduced by the large scale random
force $ f_i({\bf x},t)$  with Gaussian statistics with zero mean and
matrix of the correlation functions $D_{ij}\equiv \langle f_if_j\rangle$,
which will be specified later.

We shall consider the weakly compressible fluid when
the fields of the density and pressure can be written as  sums of
the mean values $\overline{\rho}$, $\overline{p}$ and small fluctuations
$\rho$, $p$: $\varrho = \overline{\rho} + \rho $, ${\cal P}=\overline{p}+ p$.
Without loss of generality, we take $\overline{\rho} =1$. Due to the
smallness of the fluctuations, the equation of state can be taken in the
adiabatic approximation:
\begin{equation}
 p={c^2}\rho\,,
\label{ES}
\end{equation}
where $c$ is the adiabatic speed of sound in the turbulent medium.
In the incompressibility limit one has $c^{2}=\infty$ or,
equivalently, ${\bf Ma}=0$.

Using the adiabatic relation (\ref {ES}) the continuity equation
(\ref {NR}) can be rewritten in the form
\begin{equation}
\frac 1{c^2}\partial _t p +  \frac 1{c^2}\partial_i (p v_i)
+\partial _i v_i=0.
\label{NRP}
\end{equation}
For $c^2=\infty$ the density  becomes a constant,
the velocity becomes
transversal ($\partial_i v_i=0$), and we return  to the case of
incompressible fluid.

The velocity field $v_i$ can be expressed in the form
$v_i=v_i^{\perp}+v_i^{\parallel}$,
where $v_i^{\perp}\equiv P^{\perp}_{ij}v_j$ is the transversal
part satisfying the condition $\partial_i v_i^{\perp}=0$,
and $v_i^{\parallel}\equiv P^{\parallel}_{ij}v_j$ is the longitudinal part.
The longitudinal $P^{\parallel}$ and the transversal $P^{\perp}$ projectors
in the wave-vector (${\bf k}$) space have the form
$P^{\parallel}_{ij}= k_i k_j/k^2$ and
$P^{\perp}_{ij}=\delta_{ik} - P^{\parallel}_{ij}$,  respectively
$(k\equiv\mid {\bf k}\mid).$

Since the velocity field has to become transversal in the
incompressibility limit,
its longitudinal part ${\bf v}^{\parallel}$ has to be proportional
to the inverse square of the sound speed,
$|{\bf v}| ^{\parallel}\sim c^{-2}$.
Then from the Navier-Stokes equation (\ref{NS}) it follows
that $|{\bf f}|^{\parallel}\sim c^{-2}$ for the longitudinal
part of the random force.
Hence, $c^{-2}$ can be treated as a small formal parameter, and
the compressibility corrections to the transversal part of the velocity field
can be studied within the expansion in $c^{-2}$ (or in ${\bf Ma}^2$).

In the first order in $c^{-2}$ the continuity equation (\ref{NRP})
takes the form
\begin{equation}
\frac{1}{c^2}(\partial_t  +v_i^{\perp} \partial_i) p +
\partial_i v_i^{\parallel}=0.
\label{NRPP}
\end{equation}
In the leading approximation in $c^{-2}$ (corresponding to the
incompressible fluid) the Navier-Stokes equation (\ref{NS})
gives the well-known relation between the pressure $p$ and the transversal
velocity
\begin{equation}
\Delta p = - \partial_i \partial_j v_i^{\perp} v_j^{\perp}.
\label{NSS}
\end{equation}

The last two equations allow us to express the pressure and the longitudinal
part of the velocity via the transversal part ${\bf v}^{\perp}$
\begin{equation}
 v_i^{\parallel}= - \frac{1}{c^2}\partial_i \Delta^{-1}\nabla _t p,\,\qquad
p = -\Delta ^{-1}\partial _i\partial _j v_i^{\perp}v_j^{\perp},
\label{CO}
\end{equation}
where $\nabla_t \equiv \partial_t +v_i^{\perp}\partial_i$ is the Lagrangian
derivative for the transversal part of the velocity and $\Delta^{-1}$
is the Green function for the Laplace operator. In the field theory
the quantities like the right-hand sides of Eqs. (\ref{CO})
are termed ``composite operators''.

Operating with the transversal projector $P^{\perp}$ onto Eq. (\ref{NS})
and using relations (\ref{CO}), we arrive at the closed equation for the
transversal part of the velocity (and, therefore, for all its statistical
moments) in which the compressibility is taken into account to the order
of $c^{-2}$, or, equivalently, ${\bf Ma}^{2}$
\begin{equation}
\partial_t  v_i^{\perp} = \nu_0 \Delta v_i^{\perp}
-P^{\perp}_{ij}[v_s^{\perp} \partial_s v_j^{\perp}]
-P^{\perp}_{ij}[v_s^{\perp} \partial_s v_j^{\parallel} +
v_s^{\parallel} \partial_s v_j^{\perp}]
-c^{-2}\nu_0 P^{\perp}_{ij}[p v_j^{\perp}] + f_i^{\perp}.
\label{NSC}
\end{equation}
To simplify the notation, we shall write $ v_i$ instead of $ v_i^{\perp}$
in what follows.

The positively definite $(d \times d)-$ square matrix of the
pair correlation functions  of the random force $f^{\perp}_i$ will
be taken in the form (see, e.g., \cite{Dominicis,UFN})
 \begin{equation}
 \langle f_j^{\perp}({\bf x},t)\, f_s^{\perp}({\bf 0},0) \rangle \equiv
 \varepsilon_0 D_{js}\left({\vec x}, t\right) =
 \delta(t)\,\varepsilon_0 \int\frac{{\mbox d}^d {\bf k}}{(2\pi)^d}\,
 D_{js}^{st}({\bf k})
 \exp \left[{\rm i} {\bf  k}{\bf x}  \right]\,,
 \label{CorL1}
 \end{equation}
\begin{equation}
D_{ij}({\bf k}) = k^{4-d-\epsilon} {\cal P}^{\perp}_{ij}({\bf k})
\label{cor}
\end{equation}
(we recall that $\langle f_j^{\perp} \rangle = 0$).
We see that temporal correlations of $f_i^{\perp}$ have the character
of white noise, while the spatial falloff of the correlations is
controlled by the parameter $\epsilon$ and space dimension $d$.
The functions (\ref{CorL1})  are translation invariant and for
$\epsilon=2$ become scale invariant when
the amplitude
$\varepsilon_0$ acquires the dimension of the energy dissipation rate,
$\varepsilon$, see Ref. \cite{Pismak}.
The value $\epsilon=2$ is
physically most acceptable,
since it represents the assumption that random force acts at very
large scales, which substitutes for effect of boundary conditions.
For simplicity, we use the force correlation function (\ref{CorL1})
without the usual infrared regularization. In this case, for
 $\epsilon=2$ the function (\ref{cor}) with the proper choice of
the amplitude
$\varepsilon_0$ in Eq. (\ref{CorL1}) can be considered as a power-like model
of the ``ideal'' pumping function $\delta({\bf  k})$, see \cite{UFN}.
The justification of this choice as well as the discussion of the central
problem of the $\epsilon-$expansion, i.e. the continuation from
$\epsilon=0$ to $\epsilon=2$ have been thoroughly discussed in
Ref. \cite{Antonov}.

The ratio $\varepsilon_0/\nu^3 \equiv g_0$ plays the role of a bare
coupling constant, i.e. the expansion parameter in the
nonlinearity $({\bf v}{\bf \partial}){\bf v}$ in the non-renormalized
perturbation theory. In the limit $\epsilon \rightarrow 0$ the
constant $ g_0$ becomes dimensionless,
the diagrams of the Green functions become divergent in the ultraviolet (UV)
region of the wave-vector space, and the problem of eliminating these
divergences emerges. In the field theory this problem is solved by the
well-known UV renormalization procedure, see, e.g., \cite{Collins}.

In this paper we consider the uniaxial anisotropic turbulence.
The transverse projector ${\cal P}^{\perp}$ for the correlation matrix
(\ref{cor}) is defined by the relations \cite{Barton,Kim,Busa}:
\begin{eqnarray}
{\cal P}_{js}^{\perp}({\bf  k})&=&(1+\alpha_1 \xi^2)
P_{js}^{\perp}({\bf  k}) + \alpha_2 R_{js}^{\perp}({\bf  k})\,,
\label{aniz1}\\
P_{js}^{\perp}({\bf k})&=&\delta_{js}-P_{js}^{\parallel}({\bf k})\,,
\quad
P_{js}^{\parallel}({\bf  k})= {k_j\,k_s} k^{-2}\,,
\nonumber
\\
R_{js}(\bf k) &=&( n_j - \xi_k k^{-1} k_j  )\,
                  ( n_s - \xi_k k^{-1} k_s  )\,,
\quad
\xi_k=({\bf k}{\bf n})\,k^{-1}\,,
\label{aniz}
\end{eqnarray}
where the unit vector
${\bf n}$ yields the direction of the anisotropy axis and
$\alpha_1$, $\alpha_2$ are free amplitudes. These amplitudes are not
considered small in the present analysis, however, restrictions to
their values $\alpha_1\ge -1$, $\alpha_2\ge 0$ follow from positive
definitness of the matrix (\ref{CorL1}).
For nozero $\alpha_1, \alpha_2$ the random forcing describes differences in
energy injection in the prefferred direction and directions perpendicular
to it with the subsequent generation of anisotropic structures in
large-scale eddies.

\section{Field theoretic formulation and the RG equation} \label {sec:3}

As in the critical dynamics \cite{Jansen}, the stochastic problem
(\ref{NSC}),(\ref{CorL1}),(\ref{cor}) is mapped to a quantum-field model,
which is determined by an effective De Dominicis-Janssen "action"
S({\bf v}, {\bf v}') constructed on the basis of the original
stochastic model. This action is a functional of the
transversal velocity  ${\bf v}$  and an independent transverse
auxiliary field ${\bf v'}$.

In this approach, the generating functional $G$ of the velocity
correlation and response functions is the functional integral
\begin{equation}
G({\bf A},{\bf A}')=\int D{\bf v}\,D{\bf v}' \det M ({\bf v})
\exp\left[ S({\bf v}, {\bf v}')
+ {\bf A}{\bf v}+{\bf A}'{\bf v}'\right] \,,
\label{GF}
\end{equation}
with the effective action
\begin{equation}
S({\bf v}, {\bf v}')=\frac {1}{2} g_0 \nu_0^3 {\bf v}'D{\bf v}'+
{\bf v}'[ - \partial _t{\bf v}+\nu_0 \Delta {\bf v}
 - ({\bf v}{\bf \partial}){\bf v}-
({\bf v}^{\parallel}{\bf \partial}){\bf v}-
 ({\bf v}{\bf \partial}){\bf v}^{\parallel}
- c^{-2}\nu_0 \Delta {\bf v}p],
\label{EQ}
\end{equation}
where
${\bf A}$, ${\bf A}'$ are the source fields, which are equivalent
to regular external forces. Here, the required integrations over
the spacetime arguments of the fields and sums over discrete indices
are implicitly assumed.

The Jacobian $\det M$ in Eq. (\ref{GF}) ensures the cancellation
of all the diagrams containing the self-contracted bare propagator
$ \langle {\bf v}{\bf v}'\rangle$,  which arise along with other diagrams
from the rules of the Feynman diagrammatic technique for the action
(\ref{EQ}), but do not arise in the construction of diagrams by direct iteration
of the stochastic equation (\ref{NSC}). Following \cite{Jansen,Dominicis,Pismak},
we simply define these superfluos diagrams as zero, and simultaneously
set $\det M=1$ in Eq.(\ref{GF}). We note that in our model such a definition
is nontrivial because the interaction in (\ref{EQ}) involves the derivatives
with respect to the time variable. Nevertheless, this definition is
feasible, as it has been shown in Ref. \cite{Loran} using
isotropic turbulence as an example.
As a result, we arrive at a standard field
theoretic model with action (\ref{EQ}), and the standard renormalization
theory is applicable to it.

The action (\ref{EQ}) is not renormalized and the corresponding
Green functions of the fields ${\bf v}, {\bf v}'$
contain UV divergences for $\epsilon\rightarrow 0$. In order to
analyze them, we rewrite the action (\ref{EQ}) as a sum $S=S^I+S^C$:
\begin{equation}
S^I({\bf v}, {\bf v}')=\frac{1}{2}g_0 \nu_0^3 {\bf v}'D{\bf v}'+
{\bf v}'[ - \partial _t{\bf v}+\nu_0 \Delta {\bf v}
 - ({\bf v}{\bf \partial}) {\bf v}],
\label{EQI}
\end{equation}
\begin{equation}
S^C({\bf v}, {\bf v}')=
a_{01} F_1 + a_{02}  F_2,
\label{EQC}
\end{equation}
where $a_{01}\equiv -c^{-2}, a_{02}\equiv c^{-2}\nu_0 $,
and the composite operators $F_1, F_2$, according to  Eq. (\ref{CO})
and using the relation
$\partial_i v_j^{\parallel}=\partial_j v_i^{\parallel}$, can be
represented in the form
\begin{equation}
F_1 = v_i^{\parallel}= v_i'(\partial_j v_i - \partial_i v_j)
\partial_l \Delta^{-1}\nabla _t \Delta^{-1}\partial_i
\partial _j v_iv_j,\qquad
F_2 = v_l' ( \Delta v_l) \Delta^{-1}\partial_i\partial_j v_iv_j.
\label{COV}
\end{equation}
In the limit $c^{-2}\to 0$, the action (\ref{EQI}) describes
the incompressible anisotropic turbulence.  Renormalization of this
model has been considered in \cite{Kim,Busa}. It was shown that in order
to ensure
the multiplicative renormalizability, the model has to be extended by adding
certain anisotropic dissipative terms with new viscosity coefficients
$\nu_0\chi_{0i},$ $i=1,2,3$, where the dimensionless parameters $\chi_{0i}$
describe the relative impact of the different anisotropic structures on
the viscous dissipation and play the role of additional coupling constants.

The renormalized action corresponding to the original
non-renormalized functional (\ref{EQI}) is of the form
\begin{eqnarray}
S&=&\frac{1}{2}g \nu^3 \mu^{2\epsilon}
{\bf v}'D{\bf v}'+{\bf v}'[- \partial_t{\bf v}+\nu Z_{\nu }
\Delta {\bf v}+\nu Z_{\nu }\chi_1Z_{\chi_1}{\bf n}\Delta
({\bf v}{\bf n}) +
\nonumber
\\
&+&\nu Z_{\nu }\chi_2 Z_{\chi_2}({\bf n}{\bf \partial })^2{\bf v}
 +\nu Z_{\nu }\chi_3 Z_{\chi_3}{\bf n}({\bf n}{\bf \partial})^2
({\bf v}{\bf n}) - ({\bf v}{\bf \partial}){\bf v}].
\label{ACT}
\end{eqnarray}
Here, the renormalization mass $\mu$ is an additional arbitrary parameter
of the renormalized theory, the renormalized parameters
 $g$, $\nu $, $\chi_i$ are related to their bare (unrenormalized)
 counterparts by the
multiplicative renormalization formulae, \cite{Busa}
\begin{equation}
g_0=g Z_g \mu^{2\epsilon}, \quad \nu_0=\nu Z_{\nu}, \quad
\chi_{0i}=\chi_{i}Z_{\chi_i}, \quad Z_g=Z_{\nu}^{-3}.
\label{Z}
\end{equation}
The renormalization constants $Z$ are calculated
within the perturbation theory.
In the minimal subtraction  scheme they have the form
``$Z$= 1+ only poles in $\epsilon$'' and cancel all the UV divergences
in the correlation functions of the primary fields in the model
 (\ref{ACT}). The last relation in (\ref{Z}) follows from the absence
of the constants  $Z$  in the first and  last terms of the action
(\ref{ACT}).

To determine the dependence of the renormalized correlation
functions on the parameters $a_{01}$ and $a_{02} $ after the term
(\ref{EQC}) has been added to the action, let us consider the pair
correlation function for the
incompressible isotropic case (the detailed discussion can be found,
e.g., in Refs. \cite{UFN,Antonov})
 \begin{equation}
 \langle\,
 v_{j}\left({\bf x}_1,t\right)
 v_{m}\left({\bf x}_2,t\right)
 \rangle
 =
 \int
 \frac{ {\mbox{d}}^d {\bf k} }{(2\pi)^d}\,
 G^{R}_{jm}({\bf k})\,
 \exp\left[\,{\rm i}{\bf k} \left({\bf x}_1
 -{\bf x}_2\right)\,\right]\,.
 \end{equation}

The RG equation for the trace of its space Fourier transform
$G^R({\bf k})= G^R_{ii}({\bf k})$ is
\begin{equation}
\left[
 \mu \frac{\partial }{\partial \mu}
 + \beta_{g}
 \frac{\partial }{\partial g}
- \gamma_{\nu}
 \,\,
 \nu \frac{\partial }{\partial \nu}
 \right]\,
 G^{R}({\bf k})=0\,,
\label{ERG}
 \end{equation}
where the $\beta_g$ function and the anomalous dimension
 $\gamma_{\nu}$ are expressed via the renormalization constant
 $Z_{\nu}$
\begin{equation}
\beta_g=-g (2\epsilon -3 \gamma_{\nu}), \qquad
\gamma_{\nu}= \tilde{\cal D} _{\mu}    \ln Z_{\nu}\,.
\label{B}
\end{equation}
Here $\tilde{\cal D}_{\mu}$  denotes the operation $ \mu\partial /\partial\mu $
taken at fixed values of all the bare parameters.

The solution of the RG equations along with the dimensionality
considerations gives
 \begin{equation}
 \label{ERGS}
 G^{R}({\bf k})=
  \bar{\nu}^2(s) k^{2-d}\,
 R({\overline {g}}(s)),\qquad  s\equiv\frac{k}{\mu},
 \end{equation}
where $R$ is a ``scaling function''
of the invariant charge
$\bar{g}(s)$, the effective variable
satisfying the equations
 \begin{equation}
 s\frac{\mbox{d}{\overline g} }{
 \mbox{d}s }= \beta_{g} \left( {\overline g}(s)
 \right), \qquad
 {\overline  g}|_{s=1}= g\,.
 \label{GM}
 \end{equation}
The second effective variable, the invariant viscosity $\bar{\nu}(s)$,
satisfies the equations
 \begin{equation}
 s\frac{\mbox{d}{\overline \nu} }{
 \mbox{d}s }= -\gamma_{\nu} \left( {\overline g}(s)
 \right)\,,
 {\overline  \nu}|_{s=1}= \nu\,.
 \label{GML}
 \end{equation}
From the solution of  equations (\ref{GM}) it follows that
$\overline{ g}(s)\rightarrow g^*$ for $s \rightarrow 0$, where $g^*$
is an infrared stable fixed point of the RG equations, i.e. the root of the
equation $\beta_g=0$ with the positive value of the correction exponent
$\omega \equiv  \partial \beta_g/\partial g$.

The solution of  equation (\ref{GML}) is
\begin{equation}
 {\overline  \nu} (s)=
\nu
\exp\left[-\int_g^{\overline{g}(s)}\,
 {\mbox{d} x}
 \frac{\gamma_{\nu}(x)}{\beta_g (x)}
\right]\, .
 \label{DoubW}
\end{equation}
From (\ref{DoubW}) along with Eqs.
  (\ref{B}),  (\ref{Z})  it follows that
\begin{equation}
\overline{\nu}(s)=
\nu \left( \frac{g}{\overline{g} s^{2\epsilon}}\right)^{1/3}=
 \left(\frac{\varepsilon_{0}}{\overline{g} k^{2\epsilon}}\right)^{1/3}.
\label{CP}
\end{equation}

For the spectrum of kinetic energy $E(k)\sim k^{d-1} G^{R}({\bf k})$
in the asymptotic region $s\to0$ we obtain from Eqs. (\ref{ERGS}) and
(\ref{CP})
the expression $E(k) \sim \varepsilon^{2/3} k^{-5/3}$,
which is independent of the viscosity $\nu_0$ and corresponds to the
Kolmogorov value of the exponent.

When the anisotropic case is studied [action (\ref{ACT})], the new terms
$  \beta_{\chi_j}  {\partial } G^{R}({\bf k})/{\partial \chi_j}$
are appended to the RG equation (\ref{ERG}). The new
$\beta$  functions and the anomalous dimensions $\gamma_{\chi_{i}}$
corresponding to the new dimensionless parameters $\chi_i$
\begin{equation}
\beta_{\chi_i}=-\chi_i \gamma_{\chi_i}, \qquad
\gamma_{\chi_{i}}=\tilde{\cal D}_{\mu} \ln Z_{\chi_i}
\label{B1}
\end{equation}
are expressed via the renormalization constants $Z_{\chi_i}$ in the action
(\ref{ACT}).  The additional invariant variables $\bar{\chi}(s)$
satisfy equations like Eq. (\ref{GM}). In Ref. \cite{Busa}
it has been shown that those equations have an IR stable fixed point
$\bar{g}(s), \bar{\chi}(s)$ $\rightarrow$ $g^*, \chi^*$, in which
all the eigenvalues of the matrix of the correction exponents
\begin{equation}
\omega_{ij}=\frac{\partial \beta_{g_i}}{\partial g_j}|_{g_i=g_i^*}\,,
\quad g_i\equiv g,\chi_1, \chi_2, \chi_3\,, \quad i,j = 0,1,2,3\,
\label{OMEGA}
\end{equation}
(to be precise, their real parts) are positive, i.e. the Kolmogorov
asymptotic regime conserves the stability against the strong anisotropy.

The problem becomes more involved if the compressibility is taken into
account. Let us suppose that we have managed to renormalize the action
(\ref{EQC}). Then, the new terms $\gamma_{a_i} {\cal D}_{a_i}
G^{R}({\bf k}) $
appear in the RG equation (\ref{ERG}), where $\gamma_{a_i}$ are the
anomalous dimensions of the renormalized parameters $a_{i}$.
In contrast with the parameters $\chi$, the renormalized  counterparts of the
parameters $a_{01}$ and $a_{02}$
have nonzero dimensions and, therefore,
the scaling function $R$ depends on the effective dimensionless variables
\begin{equation}
\overline{u_1}=k^2\overline{\nu}^2\overline{a}_1\,, \qquad \qquad
\overline{u_2}=k^2{\bar\nu}{\bar a}_2\,.
\label{a}
\end{equation}
The effective variables $\bar{a_1}(s)$ and $\bar{a_2}(s)$ satisfy  equations
like Eqs. (\ref{GML}).  In the infrared asymptotic region
 $k\rightarrow 0$ they take on the form
$\bar{a_i}\sim k^{-\gamma_{a_i}}$, and the infrared asymptotic form of
the dimensionless arguments (\ref{a})  is given by the expressions
$ \bar u_{i}  \sim k^{-\Delta_{a_{i}}} $ with the scaling dimensions
$\Delta_{a_i}$(for more details see, e.g.,
\cite{UFN,Rel,Vass}).
In the linear approximation with respect to the small parameters
$a_{01} and a_{02}$ in the functional (\ref{GF}) the leading correction
to the scaling function  $R$ takes on the form
$ (1+ {\rm const}\cdot k^{-\Delta_{max}}) $ where
$\Delta_{max}$ is the maximal dimension among $\Delta_{a_{i}}$.

Therefore, the investigation of the dependence of the
kinetic energy spectrum
on the compressibility is related to the calculation of the scaling
dimensions $\Delta_{a_{i}}$ which, as we shall shown below, can be
expressed via scaling dimensions of the composite operators $ F_{1,2}$
entering into the action (\ref{EQC}).

\section{Renormalization and scaling dimensions of the composite
operators}  \label {sec:4}

The addition of the term (\ref{EQC}) involving the operators
$F_1, F_2$ (\ref{COV}) to the action (\ref{EQI}) gives rise to new
UV divergences (poles in $\epsilon$) in the correlation functions.
According to the generic rules,  all the composite operators with
the same canonical (naive) dimensions and tensor structure can be mixed in
the renormalization procedure, i.e. an UV finite renormalized operator
$F^R$ has the form $F^R=F+counter terms,$ where the contribution
of the counter
terms is a linear combination of $F$ itself and other unrenormalized
operators that "admix" to $F$.
Therefore, to perform renormalization of the operators $F_1 and F_2$,
one has to consider a
wider family of operators $F_i$ which admix to $F_1$, $F_2$.

The renormalized operators $F^R_i$ are related to their
non-renormalized counterparts $F_i$ by the well-known matrix
formulae of multiplicative
renormalization, see, e.g. Refs.  \cite{Pismak,UFN}
\begin{equation}
F_i=Z_{ij}F_j^R,
\label{OMZ}
\end{equation}
where  $Z_{ij}$ is the matrix of the renormalization constants.
In the minimal subtraction scheme its diagonal elements have the form
$1 + $ poles in $\epsilon$ while the non-diagonal elements
contain only poles.
From the matrix  $Z_{ij}$ one calculates
the matrix of anomalous dimensions
$\gamma_{ij}=Z^{-1}_{ik}\tilde {\cal D}_{\mu} Z_{kj}$ and the matrix
of scaling dimensions for the
set of operators
\begin{equation}
\Delta_{ij}^F=D^F_{ij}+\gamma_{ij}.
\label{CD}
\end{equation}
The contribution $D^F_{ij} = [d_F - \gamma_{\nu}
d^{\omega}_F]_{ij}$ is expressed via the anomalous dimension of the viscosity
(\ref{B}), and the total
 $d_F$ and frequency $d^{\omega}_F$ canonical dimensions of the operator
  $F$ \cite{Pismak,UFN}, which are equal to the sums of  corresponding
dimensions of the fields
and derivatives that constitute  $F$.

The total canonical dimensions of the fields and parameters of the model
are found from the requirement that all the terms of the action (\ref{EQ})
be dimensionless, see \cite{Pismak,UFN}:
$d_t= d_{\omega}=-2$, $d_v=1,$ $d_{v'}=d-1$ ($d_{ x}
= d_{k}=-1$ by definition).
From these dimensions we then obtain the canonical
dimensions of the operators $F_1$, $F_2$ equal to $d_{F_1}=d_{F_2}=d+4$.
We also note,
that these operators are Galilean invariant, scalar and nonlocal.

Let $F\equiv\{F_{i}\}$ be a system of composite operators closed
with respect to renormalization. The equation $a_iF^{R}_i=a_{0i}F$
(the summation over the subscript $i$ is implied)
can be regarded as a definition of the renormalized sources
$a\equiv\{a_{i}\}$, which for the usual renormalization formulae
$a_{0}=aZ_{a}$, $F=Z_{F}F^{R}$ leads to the relations
$Z_{a}=Z_{F}^{-1}$ for the renormalization constants and
$\gamma_{F}=-\gamma_{a}$ for the corresponding anomalous dimensions.
The requirement that the terms
$$ \int dx aF^{R}=  \int dx a_{0}F \quad , x\equiv {\bf x},\, t $$
be dimensionless then gives the ``shadow relations'' for the canonical and
scaling dimensions of the operators $F_i$ and sources $a_i$
\begin{equation}
d_{a}^{k}+d_{F}^{k}=d,\quad d_{a}^{\omega}+d_{F}^{\omega}=1,\quad
\Delta_{a}+\Delta_{F} =d+\Delta_{\omega}.
\label{shadows}
\end{equation}
Due to Eqs. (\ref{shadows}), the problem of finding the maximal
dimension $\Delta_{a_{i}}$
for the sources corresponding to the operators $ F_{1,2}$ in the action
(\ref{EQC}) is equivalent to the calculation of the minimal
scaling dimension $\Delta_{F} $ associated with the operators $ F_{1,2}$
and all the operators that admix to them in renormalization.

According to the general theory of renormalization,
see, e.g. \cite{Collins}, counter terms in a field theory with a local
interaction are also local. Therefore, the renormalization of the nonlocal
operators $F_1$, $F_2$ is reduced to that of their local
blocks (see below) and to the admixture of the local operators
(i.e. monomials constructed of the fields and their derivatives
at the same point ${\bf x}$, $t$) with the same canonical dimension
and symmetry
(Galilean invariant scalars). These local operators in our case are
the following:
$\overline{F} = \partial v' \partial v \partial v$,
$\partial v' \nabla_t v \partial v$,
$\partial v' \partial^3 v $,
$n^2\partial v' \partial v \partial v$,
$n^2\partial v' \nabla_t   \partial v$,
$n^2\partial v' \partial^3 v $,
$n^4\partial v' \partial v \partial v$,
$n^4\partial v' \nabla_t   \partial v$,
$n^4\partial v' \partial^3 v $,
$n^6\partial v' \partial v \partial v$,
$n^6\partial v' \nabla_t   \partial v$,
$n^6\partial v' \partial^3 v $.
The notation is symbolic and it implies all possible contractions
of the vector indices of the fields ${\bf v}'$, ${\bf v}$,
derivative ${\bf \partial}$ and  unit vector ${\bf n}$.
This set of operators is closed
with respect to renormalization because the nonlocal operators
$F_1$ and $F_2$ cannot admix to them. The first three types of
the operators $\overline{F}$ have been considered in \cite{Loran}. It
was shown that they did not affect the scaling dimensions of
the nonlocal operators $F_1$, $F_2$ due to the fact that the
corresponding renormalization matrix $Z_{ij}$ was block-triangular.
This feature of the renormalization matrix persists also in the
other operators $\overline{F}$, which contain the vector ${\bf n}$,
so that they also do not affect the scaling dimensions of
$F_1$, $F_2$. In contrast with the local operators $\overline{F}$,
they contain additional factors of $\Delta^{-1}\partial v$ which
have zero canonical dimension and negative scaling dimension
$-4/3$ at $\epsilon=2$ (we recall that the scaling dimension of the
field ${\bf v}$ equals to $-1/3$ at $\epsilon=2$, see \cite{Pismak,UFN}).
Therefore, the scaling dimensions of the operators $\overline{F}$ are
greater than
the dimensions of the nonlocal operators $F_1$, $F_2$,
and the leading contribution to the IR asymptotic form of the spectrum
is determined by the contributions of $F_1$ and $F_2$.
We note that due to renormalization, scaling dimension of an operator $F$
does not  coincide in general with a naive sum of scaling dimensions
of the fields and derivatives entering into $F.$
But, for the
incompressible case, the hypothesis that scaling dimension of a
nonlocal operator is the sum of scaling dimensions of its local parts and
of the factors of type $\Delta^{-1}\partial v$
has been confirmed in  \cite{Rel}
by the explicit one-loop calculation of the scaling dimensions
related to the local operators  with the canonical dimension $d+4$,
and we also accept it in what follows.

As result, we obtain that the scaling  dimensions of $F_1$ and $F_2$ are
determined by their own
renormalization. The latter is reduced to the renormalization of the local
blocks entering into $F_1$ and $F_2$.
%, cf. \cite{Loran}.

Let us denote the field ${\bf v}$ by the solid line, ${\bf v}'$ by the
oriented solid line, and the operator $\Delta^{-1}$ by the wave line.
The derivative with respect to coordinate is denoted by a slash, and the
 derivative with respect to time  by a cross.
Graphical representation of  the operators
(\ref{COV}) is depicted in Fig.~1, where the vector
indices are omitted and the operator, containing the full time derivative
$\nabla_t$, is represented as a sum of the first two diagrams.

The contribution of the last operator from Fig.~1 to the correlation
function $\langle {\bf v}'{\bf v}{\bf v}{\bf v}\rangle$ is depicted
in Fig.~2. The shadowed rectangle denotes an
arbitrary one-particle irreducible diagram with fixed external legs.
One can show that the triangular subdiagram contains  UV divergence and
its elimination requires the renormalization of the local block
of the nonlocal operator under consideration.
Thus, for the complete
renormalization  of the operators (\ref{COV}) it is sufficient to
study the renormalization  of all their local blocks.

The operator $F_1$ consists of two nonlocal factors $\Delta ^{-1}$,
the full derivative $\nabla_t $, and two local blocks
\begin{equation}
G_1= v_i'(\partial _jv_i-
\partial _iv_j), \qquad
G_2=\partial_i\partial_j v_i v_j     ,
\label{OP2}
\end{equation}
while $F_2$ contains one factor $\Delta ^{-1}$, the operator $G_2$,
and the local block
\begin{equation}
G_3= v_i'\Delta v_i.
\label{OP3}
\end{equation}
The scaling dimensions of the operators (\ref{COV}) are equal to the sums
of the scaling dimensions of the above factors, among them only the
dimensions
of the local blocks (\ref{OP2}) and (\ref{OP3}) require nontrivial
calculation. In order to find them one has to study the renormalization
of the complete set of the operators that admix to $G_i$
in renormalization.
This set is rather big because of the anisotropy
and the canonical dimension of $G_i$ is high ($d_F=7$ for $d=3$).
To simplify the analysis, we shall use some general rules
for the operator mixing. Their proof and other examples can be found,
e.g., in Refs. \cite{Pismak,UFN,Hnatic,55}.

\begin{description}

\item[(a)] In the action  (\ref{ACT}) the derivative in the
interaction term can be moved onto the auxiliary field ${\bf v}'$
using integration by parts:
$v'_iv_j\partial _jv_i=-(\partial _jv'_i)v_iv_j$.
Therefore, the derivative $\partial$ appears as an external factor for
each external leg of the field ${\bf v}'$ for any one-particle irreducible
diagram, and the corresponding counter term contains the factor
${\bf \partial} {\bf v}'$.

\item[(b)] Only Galilean invariant operators can admix to an invariant
operator in the renormalization procedure.

\item[(c)] Let some operator $G$ has the form of a total derivative
of some other operator $[G]$,
$G=\partial [G]$. In this case, the scaling
dimension of $G$ is simply given by the relation
$\Delta _G=1+\Delta _{[G]}$.

\item[(d)] All the one-particle irreducible diagrams,
containing closed circuits
of the retarded propagators $\langle{\bf v}{\bf v}'\rangle$, vanish.
\end{description}

We denote by $\tilde{G}$ or $[\tilde{G}]$ the full sets of operators
that can mix with a given $G$ or $[G]$ in renormalization.

According to the item (c), instead of the operator $ G_2$ from (\ref{OP3})
it is sufficient to study the renormalization of the operator
$[G_2]=v_i v_j$. Due to the transversality of the field
$v_i$ the only operators that can admix to $[G_2]$ have the form
$[\tilde{G}_2]=n_kn_l\partial _lv_k\delta _{ij}$,
$n_k n_l\partial_lv_k n_i n_j$.
Their scaling dimensions are equal to $\Delta_{[\tilde{G}_2]}=1+\Delta_v$.
The scaling dimensions of the fields ${\bf v}$, ${\bf v}'$ and the time
have the form (see, e.g. \cite{Pismak})
\begin{equation}
\Delta_v = 1- 2\epsilon/3, \quad
\Delta_{v'} = d-1 + 2\epsilon/3, \quad
\Delta_{t} = -2 + 2\epsilon/3.
\end{equation}
We then obtain $\Delta_{[\tilde{G}_2]}=2 - 2\epsilon/3$,
which gives $\Delta_{[\tilde{G_2}]}=2/3$ at $\epsilon=2$.
 Since the operator $[G_2]$ itself is not renormalized,
for the scaling dimension of $G_2$ we obtain (item (c)):
$\Delta_{G_2}=2+\Delta_{[G_2]}$$=2+2\Delta_v=4-4\epsilon/3$,
which gives $\Delta_{G_2}=4/3$  at $\epsilon=2$.

The operator $G_1 $ consists of two terms: $G_1=G_{11}-G_{12}$.
The term $G_{12}$ is rewritten in the form $G_{12}=\partial_i(v'_iv_j)$;
it is then sufficient to consider the operator $[G_{12}]=v'_i v_j$
(item (c)). It can mix with the following operators:
$\partial _iv'_j$, $n_in_l\partial _lv'_j$,
$\delta _{ij}n_kn_l\partial _lv'_k$, and $n_in_jn_kn_l\partial _lv'_k$.
They all are UV finite and  their critical dimensions are simply given by
$1+\Delta_{v'}=d+2\epsilon/3$, i.e., $13/3$ at $d=3$ and $\epsilon=2$.
The diagonal element of the matrix $Z_{ij}$ of the above set of operators
equals 1 (item (a)) and, as in the case of the set associated with
$G_2$, this matrix is triangular. It then follows that
$\Delta_{G_{12}}=1+\Delta_{[G_{12}]}=1+\Delta_v+\Delta_{v'}=
d+1$, which gives $\Delta_{G_{12}}=4$ at $d=3$.

The operator $G_{11}$ does not admix to itself due to the item (a).
Owing to the Galilean invariance, it does not mix with the operators
of the same tensor structure which involve the vector ${\bf n}$ (item (b)).
Furthermore, it does not mix with the invariant operators
 $n_j\nabla _tn_kv'_k$ (item (a)) and $n_i\nabla _t\partial _jv_i$
(item (d)).
The set  $\tilde G_{11}$, which can admix to
$G_{11}$, includes the operators  $\Delta v'_j$, $n_j\Delta v'_s n_s$,
$ n_s\partial_s n_l\partial_l v'_j$,
$ \partial_j n_s\partial_s v'_l n_l$,  and
$ n_j n_k\partial_k n_s\partial_s v'_l n_l$.
All these operators are UV finite and their critical dimensions are
equal to $2+\Delta_{v'}=16/3$. Like the case of the operators
$G_2$ and $G_{12}$, these operators do not affect the critical
dimension of $G_{11}= v'_i\partial_j v_i$.  Since the latter is UV finite,
its critical dimension is given by
$\Delta_{G_{11}}=1+\Delta_{v'}+\Delta_{v}=4$.

Now let us turn to the last operator $G_3$ from (\ref{OP3}).
The invariant operators  $v'_i\nabla _tv_i$ and
$v'_i n_i\nabla _tv_j n_j$ do not admix to  $G_3$ due to item (a).
Therefore, we are left with the three types of operators
\begin{eqnarray}
\{\tilde{G}_{31}\}
&=& \{ v'_i({\bf n}{\bf \partial})^2 v_i, \qquad
 ({\bf n}{\bf v}')\Delta ({\bf n}{\bf v}), \qquad
({\bf n}{\bf v}')({\bf n}{\bf \partial})^2({\bf n}{\bf v})\},
\nonumber
\\
\{\tilde{G}_{32}\}
&=& \{\partial_l(v'_s \partial_l v_s),\,\,
\partial_l({\bf v}'{\bf \partial} v_l),\,\,
({\bf n}{\bf \partial})[v'_l ({\bf n}{\bf \partial}) v_l],\,\,
({\bf n}{\bf \partial})[({\bf v}'{\bf \partial})({\bf n}{\bf v})],
\,\,
\nonumber
\\
& & \partial_l [({\bf n} {\bf v}')\partial_l({\bf n}{\bf v})],\,\,
\partial_l [({\bf n} {\bf v}')({\bf n}{\bf \partial}) v_l], \,\,
({\bf n}{\bf \partial})[({\bf n} {\bf v}') ({\bf n}{\bf \partial})
({\bf n}{\bf v})]\},
\nonumber
\\
\{\tilde{G}_{33}\}
&=& \{({\bf n}{\bf \partial})\Delta ({\bf n} {\bf v}'),\qquad
({\bf n}{\bf \partial})({\bf n}{\bf \partial})^2({\bf n} {\bf v}')\}.
\label{OP31}
\end{eqnarray}
The operators $\{\tilde{G}_{33}\}$ do not affect the
scaling dimensions of
$\{\tilde{G}_{31}\}$ and $\{\tilde{G}_{32}\}$ (item (c)), they are UV finite
and their dimensions are equal to $\Delta_{\{\tilde{G}_{33}\}}=19/3$
at $d=3$ and $\epsilon=2$. The operators
$\{\tilde{G}_{32}\}$ do not affect $\{\tilde{G}_{31}\}$ (item (c)),
they are also finite (like $G_{12}$), and their scaling
dimensions are equal to $\Delta{\{\tilde{G}_{32}\}}=5$.

Thus, we need to renormalize  the remaining set
that includes the operators $G_3$ and $\tilde{G}_{31}$. They are
renormalized with mixing, and the corresponding matrix $Z_{ij}$
is nontrivial. In isotropic case
the renormalization constant of $G_3$ is expressed via the known
renormalization constant $Z_{\nu}$ in the action (\ref{ACT}) and, therefore,
the scaling dimension $\Delta_{G_3}$ is related to the known
function $\gamma_{\nu}$ \cite{Loran}.
In the presence of anisotropy the situation
becomes more complicated. However, even in this case it turns out
possible to express the matrix $Z_{ij}$ in terms of the known
renormalization constants $Z_{\nu}$ and $Z_{\chi_i}$ from the action
(\ref{ACT}).

Consider the generating functional (\ref{GF}) with $\det M = 1$ and the
renormalized action (\ref{ACT}).  It is UV finite and, therefore,
its derivative with respect to the renormalized parameters
 $e=\{ g, \chi_i, \nu \} $  (they are the generating functionals
of the composite operators $\partial _e S$)
are also UV finite, as well as the operators $\partial _e S$ themselves.

The functional $G({\bf A}, {\bf A}')$ satisfies the RG equation
\begin{equation}
{\cal D}_{RG}G({\bf A}, {\bf A}')=0,\qquad {\cal D}_{RG}=[\mu
\frac{\partial}{\partial \mu}-
\gamma_{\nu }\nu\frac{\partial}{\partial {\nu}}+\beta_g
\frac{\partial}{\partial g}+\beta _{\chi_i}
\frac{\partial}{\partial {\chi_i}}]
\label{RGG}
\end{equation}
with the functions $\beta_g$ and $\beta_{\chi_i}$ defined in
 (\ref{B}) and (\ref{B1}).
Let us define the matrix  $\omega_{ik}$ by the relation
\begin{equation}
\omega_{ik}
=-g_i \frac {\partial \gamma_{g_i}}{\partial g_k},
\label{OM}
\end{equation}
where $g_i\equiv \{g, \chi_i\} $ (we recall that $\gamma_g=-3\gamma _{\nu}$).
Using (\ref{B}) and (\ref{B1}) we readily find that at the fixed point
$g_i^*$ the matrix (\ref{OM}) coincides with the matrix of correction
exponents $\omega _{ik}$ defined in (\ref{OMEGA}).

We define the commutator of two differential operators ${\cal D}_i$,
${\cal D}_j$ in a standard way, $[{\cal D}_i, {\cal D}_j]\equiv {\cal D}_i
{\cal D}_j - {\cal D}_j {\cal D}_i$.
The commutators of the operators $\cal D_{RG}$, ${\cal D_{\nu}} \equiv
\nu \partial_{\nu} \equiv \nu
\partial/\partial \nu $ and
$\partial_{g_i} \equiv \partial/ \partial g_{i}$ are of the form
\begin{equation}
[{\cal D}_{RG}, {\cal D}_{\nu}]=0, \quad
[{\cal D}_{RG}, \partial_{g_i}]=\omega_{ij}
[\delta_{i0} \frac{1}{3g}
{\cal D}_{\nu}-\
\partial_{g_j}] - \delta_{i0} \frac{\beta_g}{g} \partial_g  .
\label{CR}
\end{equation}
Differentiation of the RG equation (\ref{RGG}) with respect to ${\nu }$
and ${g_i}$ along with the commutation relations (\ref{CR}) gives
\begin{equation}
{\cal D}_{RG}\partial_{g_i} G=\omega_{ij}
[\delta_{i0} \frac 1{3g}
{\cal D}_{\nu}-\partial _{g_j}]G
- \delta_{i0} \frac{\beta_g}{g} \partial_g G, \qquad
{\cal D}_{RG}{\cal D}_{\nu}G=0.
\label{SYS}
\end{equation}
The fact that the operators ${\cal D}_{RG}$ and ${\cal D}_{\nu}$ are
commutative allows the left-hand side of the first equation in (\ref{SYS})
to be rewritten in the form
$${\cal D}_{RG}\partial_{g_i} G = {\cal D}_{RG}
[\partial_{g_i} - \delta_{i0} \frac 1{3g} {\cal D}_{\nu}]G-\delta_{i0}
\frac{\beta_g}{3g^2}{\cal D}_{\nu} G. $$
Using this relation,  equation (\ref{SYS}) is rewritten as
\begin{equation}
{\cal D}_{RG}X_i=-\omega _{ij}X_j - \delta_{i0}\frac{\beta_g}{g}
X_0\,, \qquad i,j = 0,1,2,3,4
\label{MAIN12}
\end{equation}
which, at the fixed point $g^*\neq 0$ along with $\beta_g=0$, gives
\begin{equation}
{\cal D}_{RG}X_i= -\omega _{ij}X_j\, .
\label{MAIN1}
\end{equation}
This is nothing else than the scaling equation for the quantities
 $$X_i=(\partial_{g_i}-(3g)^{-1}\delta _{i0}{\cal D}_{\nu })\,
G({\bf A}, {\bf A}'),$$
and $\omega_{ij}$ is the matrix of their anomalous dimensions. Its
eigenvalues are positive (it follows from the IR stability of the
fixed point, see \cite{Busa}). According to (\ref{OM}), it is expressed
via the renormalization constants $Z$ of the action (\ref{ACT})
calculated in the one-loop approximation in Ref. \cite{Busa}.

Using the explicit form of the generating functional (\ref{GF}),
the quantities $X_i$ are explicitly expressed via the derivatives
of the renormalized action (\ref{ACT}) with respect to the parameters
$g$, $\nu$, and $\chi_i$
\begin{eqnarray}
X_i =
\int D{\bf v} D{\bf v}'
\tilde{X}_i \exp [S({\bf v}, {\bf v}') +{\bf A} {\bf v} +{\bf A}'{\bf v}'] =
\nonumber
\\
=\int D{\bf v}D{\bf v}'
[\partial_{g_i} S-(3g)^{-1}\delta _{i0}{\cal D}_{\nu } S]
\exp [S({\bf v}, {\bf v}') +{\bf A} {\bf v} + {\bf A}' {\bf v}'] .
\label{MAINS}
\end{eqnarray}
Therefore, the quantities $X_i$ represent the generating functionals
of the correlation functions that involve not only primary fields
${\bf v}$, ${\bf v}'$ but also renormalized composite operators
$\tilde{X}_i$.

Performing the differentiation in (\ref{MAINS}) explicitly one obtains
\begin{equation}
\tilde{X}_i=e_{i0} {\bf v}'\Delta {\bf  v}+
e_{i1}({\bf  v}'{\bf  n})\Delta ({\bf  v}{\bf  n})
+e_{i2}{\bf  v}'({\bf  n}{\bf  \partial})^2{\bf  v}
e_{i3}({\bf  v}'{\bf  n})({\bf  n}{\bf  \partial})^2({\bf  v}{\bf  n}),
\label{OPE}
\end{equation}
where the coefficients $e$ are expressed via the renormalization constants
from (\ref{ACT}):
\begin{eqnarray}
e_{00}& = & \nu\left(\partial_g  Z_{\nu }-
\frac{Z_{\nu}}{3 g}\right)\mid_{{ g}={ g}^*}\,,\qquad
e_{0i} =\nu\chi_i\left[\partial_g (Z_{\nu }  Z_{\chi_i})-
\frac{Z_{\nu }Z_{\chi_i}}{3 g}\right]\mid_{{ g}={ g}^*}\,,
\quad i=1,2,3
\nonumber\\
e_{10} & = & \nu\partial_{\chi_1} Z_{\nu }\mid_{{ g}={ g}^*}\,,\qquad
e_{1i} =\nu\chi_i\partial_{\chi_1}
\left(Z_{\nu} Z_{\chi_i }\right)\mid_{{ g}={ g}^*}\,,
\nonumber\\
e_{20} & = & \nu\partial_{\chi_2} Z_{\nu }\mid_{{ g}={ g}^*}\,,\qquad
e_{2i} =\nu\chi_i\partial_{\chi_2}
\left(Z_{\nu } Z_{\chi_i }\right)\mid_{{ g}={ g}^*}\,,
\nonumber\\
e_{30} & = & \nu\partial_{\chi_3} Z_{\nu }\mid_{{ g}={ g}^*}\,,\qquad
e_{3i} =\nu\chi_i\partial_{\chi_3}
\left(Z_{\nu } Z_{\chi_i }\right)\mid_{{ g}={ g}^*}.
\label{FIN}
\end{eqnarray}

It is obvious from Eqs. (\ref{FIN}) that  $\tilde X_i$ are given by linear
combinations of the operators $G_3$ and $\{ \tilde G_{31}\} $, and the
matrix $\omega$ (\ref{OM}) determines their anomalous dimensions.
The eigenvalues $\omega_i$ of the matrix
$\omega$ have been calculated in Ref. \cite{Busa} in the
first order of the $\epsilon$ expansion. All the real parts of these
eigenvalues are positive (two of the eigenvalues are complex).
We calculate from Eq. (\ref{CD}) the scaling dimensions of the operators
$G_3$ and $\{ \tilde G_{31}\}$
$\Delta_{G_{3}}= 13/3 + \omega$ $(\omega\equiv \omega_0),$
$\Delta_{\{\tilde G_{31}\}}= 13/3 + \omega_i$ for $i=1,2,3$.
From the results of Ref.\cite{Busa} it follows that the exponent $\omega$
is smaller than each of the eigenvalues $\omega_i$, so that the main
contribution of the operators in consideration to the IR asymptotic
behaviour of the kinetic energy spectrum is given by the operator $G_3$.

Finally, from Eqs. (\ref{COV}), (\ref{OP2}), (\ref{OP3}) and
"shadow relation" (\ref{shadows}) we obtain
the scaling dimensions for the original composite operators
$F_1$, $F_2$ and the corresponding parameters $a_1 and a_2$
\begin{eqnarray}
\Delta _{F_1} &=& d+4-2\epsilon, \qquad \Delta_{F_2}=d+4-2\epsilon+\omega,
\label{RES}\\
\Delta_{a_1} &=& 4\epsilon/3 -2,  \qquad
\Delta_{a_2} = 4\epsilon/3 -\omega.
\label{CDP}
\end{eqnarray}
For $d=3$ and $\epsilon=2$ this gives $ \Delta_{a_1} =2/3,$
$\Delta_{a_2} =2/3 -\omega$ ($\Delta_{a_2} =-10/3 $ in the first order in
$\epsilon$).

Since the parameter $a_1$ is not renormalized (see above), we have
$\overline{a}_1=a_1=a_{01}=-c^{-2}$, which along with Eqs. (\ref{a})
and (\ref{CP}) in the IR asymptotic region for the effective variable
$\overline{u}_1$ gives:
$\overline{u}_1\to u_1^*\sim c^{-2}\varepsilon^{1/3} k^{-2/3}$.
Using the well-known relation $\varepsilon \sim v_c^3/L$
(where $\varepsilon $ is the mean dissipation rate,
$v_c$ is the mean-square velocity field,
 and $L$ is the outer scale of turbulence) the expression for
$u_1^*$ can be rewritten as
\begin{equation}
u_1^*\sim ({\bf Ma})^2 (kL)^{-2/3}.
\label{u1}
\end{equation}
In a similar way, one can find the $k$ dependence of the variable
$u_2^*$ at $\epsilon = 2$. From the relation
$u_2^*\sim u_1^* (kl)^{\omega}$ (see \ref{a}),
where $ l=\varepsilon^{-1/4}\nu_0^{3/4}$ is the Kolmogorov
dissipation length, one obtains
\begin{equation}
u_2^*\sim ({\bf Ma})^2 (kL)^{-2/3} (kl)^{\omega} ,
\label{u2}
\end{equation}
and  $u_1^* >> u_2^*$ ($\omega > 0$)
in the inertial range $kl<<1.$
Therefore, the leading contribution
to the small $k$ behavior of the scaling function $R$ from Eq.(\ref{ERGS})
is given by the term with the variable $u_1^*$. In the linear approximation
in the Mach number, the leading correction to the kinetic energy spectrum
is of the form
\begin{equation}
E(k) \sim \varepsilon^{2/3} k^{-5/3}\left[1 + A
({\bf Ma})^2 (kL)^{-2/3}\right],
\label{S}
\end{equation}
where $A$ is a numerical factor. This correction is independent of the
viscosity coefficient $\nu_{0}$ which proves the validity of
the second Kolmogorov
hypothesis.
The contribution  of $u_2^*$
gives rise to a $\nu_{0}$ dependent term, but in the inertial range it only
determines a vanishing correction. For ${\bf Ma}<<1$ the correction is
rather small because in the inertial range one has $(kL)^{-2/3}\leq 1$.
In contrast with the isotropic model,
the amplitude factor in (\ref{S}) and the coefficient $A$ depend on the
anisotropy parameters.

\section{Conclusion} \label {sec:5}

We have shown that in the statistical model of the fully developed
turbulence in the presence of uniaxial anisotropy the kinetic
energy spectrum in the inertial range is independent of the
viscosity coefficient (i.e. the second Kolmogorov hypothesis holds)
in the leading approximation in the Mach number.

In this paper, we have dealt only with the dependence on the  UV scale
(or, equivalently, on the viscosity coefficient) and have not discussed
the dependence on the integral scale $L$. The RG approach along with the
operator
product expansion is also applicable to this problem. The most singular
$L$ dependence is revealed by the different-time velocity correlations
and physically is explained by the well-known sweeping effects, see, e.g.,
\cite{Reports}. The RG treatment of this problem has been given in
Ref. \cite{Antonov}
(see also Ref. \cite{UFN}) and it is readily generalized to our case.
It is now
generally accepted that the intermittency phenomenon leads to a
singular $L$ dependence of the equal-time correlations, see, e.g. Ref.
\cite{Legacy}.  In Ref. \cite{PRE},
it has been applied to the simple example of the so-called rapid-change model
of passive scalar advection \cite{RC} in order to confirm the singular
dependence of the equal-time correlation functions on $L$ and calculate
the corresponding anomalous exponents within the $\epsilon$ expansion;
the results obtained are in agreement
with the previous results obtained using the so-called zero-mode technique
\cite{zero}. The generalization of these results to more realistic models
like the stochastic Navier--Stokes requires a considerable improvement of
the existing technique and remains an open problem.
\bigskip

The work was supported by the Slovak Academy of Science (Grant No
2/4171/97), the Russian Foundation for Fundamental
Research (Grant No 96-02-17-033), and the Grant Center for
Natural Sciences of the Russian State Committee for Higher Education
(Grant No 97-0-14.1-30).

\newpage
\input epsf
   \begin{figure}[htb]
\vspace{-0.5cm}
\begin{center}
\leavevmode
\epsfxsize=7cm
\epsfbox{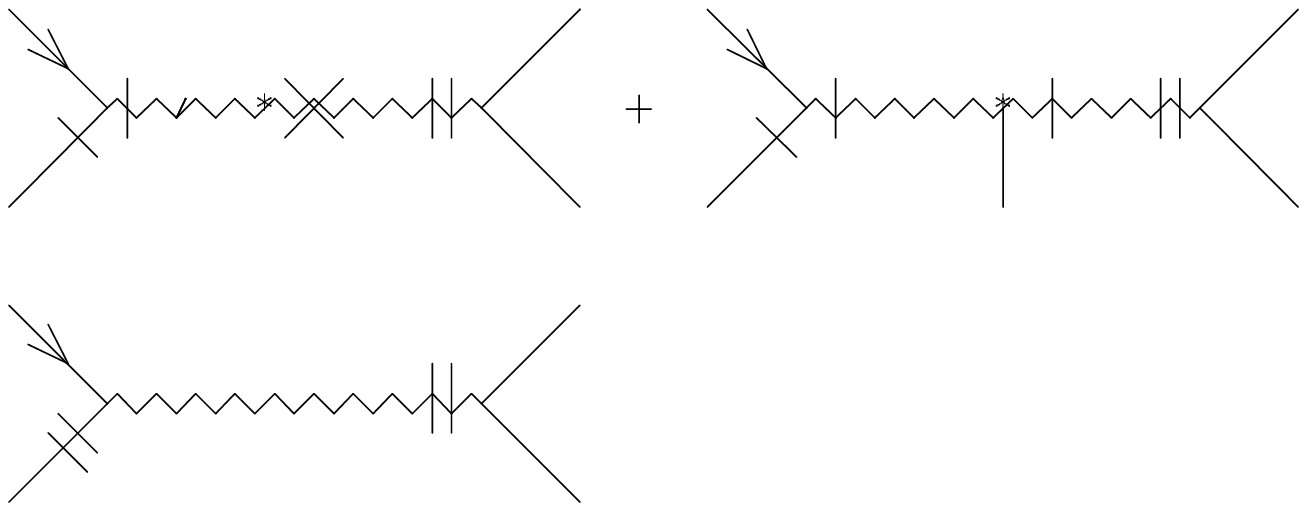}
\end{center}
\caption{
Graphs of composite nonlocal operators
$F_1 = v_i^{\parallel}= v_i'(\partial_j v_i - \partial_i v_j)
\partial_l \Delta^{-1}\nabla _t \Delta^{-1}\partial_i
\partial _j v_iv_j $ and
$F_2 = v_l' ( \Delta v_l) \Delta^{-1}\partial_i\partial_j v_iv_j$
giving a leading correction to the infrared form kinetic energy spectra of
weakly compressible developed turbulence.}
\label{1}
\end{figure}
\vspace{2cm}

\input epsf
   \begin{figure}[htb]
\vspace{-0.5cm}
\begin{center}
\leavevmode
\epsfxsize=7cm
\epsfbox{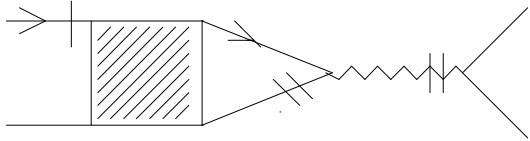}
\end{center}
\caption{
The correlation
function $\langle {\bf v}'{\bf v}{\bf v}{\bf v}\rangle$ with
the contribution of the nonlocal operator
$F_2 = v_l' ( \Delta v_l) \Delta^{-1}\partial_i\partial_j v_iv_j.$
The shadowed rectangle denotes an
arbitrary one-particle irreducible diagram with four external legs.}
\label{2}
\end{figure}
\end{document}